\documentclass[a4paper,11pt]{article}
\usepackage[dvips]{graphicx}
\usepackage{amssymb}
\usepackage{amsmath}
\usepackage{cite}

\begin{document}

\title{Dark Energy and Equation of State Oscillations with Collisional Matter Fluid in Exponential Modified Gravity }
\author{
V.K. Oikonomou$^{1}$\,\thanks{v.k.oikonomou1979@gmail.com; voiko@physics.auth.gr}\\
$^{1)}$Department of Theoretical Physics, Aristotle University of Thessaloniki,\\
54124 Thessaloniki, Greece
\\\\
N. Karagiannakis$^{2}$\\
$^{2)}$ Polytechnic School, Aristotle University of Thessaloniki,\\
54124 Thessaloniki, Greece\\\\
Miok Park$^{3}$\\
$^{3)}$School of Physics, Korea Institute for Advanced Study,\\
 Seoul 130-722, Korea
} \maketitle

\begin{abstract}
We study some aspects of cosmological evolution in a universe described by a viable curvature corrected exponential $F(R)$ gravity model, in the presence of matter fluids consisting of collisional matter and radiation. Particularly, we express the Friedmann-Robertson-Walker equations of motion in terms of parameters that are appropriate for describing the dark energy oscillations and compare the dark energy density and the dark energy equation of state parameter corresponding to collisional and non-collisional matter. In addition to these, and owing to the fact that the cosmological evolution of collisional and non-collisional matter universes, when quantified in terms of the Hubble parameter and the effective equation of states parameters, is very much alike, we further scrutinize the cosmological evolution study by extending the analysis to the study of matter perturbations in the matter domination era. We quantify this analysis in terms of the growth factor of matter perturbations, in which case the resulting picture of the cosmological evolution is clear, since collisional and non-collisional universes can be clearly distinguished. Interestingly enough, since it is known that the oscillations of the effective equation of state parameter around the phantom divide are undesirable and unwanted in $F(R)$ gravities, when these are considered for redshifts near the matter domination era and before, in the curvature corrected exponential model with collisional matter which we study here there exist oscillations that never cross the phantom divide. Therefore, this rather unwanted feature of the effective equation of state parameter is also absent in the collisional matter filled universe.  
\end{abstract}


\section*{Introduction}

Late time acceleration is one of the most profound observations made for the evolution of the universe \cite{riess}. Current experimental research \cite{planck,bicep} aims to further enlighten both the late time era and also evolution stages that belong to higher redshift. One of the latest observational successes was the verification of the B-mode power spectrum \cite{bicep}, which verifies the existence of the inflationary era \cite{mukhanov} of the universe. In addition to observational methods that use standard candles, such us Supernovas and Gamma Ray Bursts, there exist other methods proposed in the literature that may reveal the way that our universe evolves in cosmic time, using indirect methods, such us by observing the spectrum of direct dark matter scattering \cite{kinezosvergados}. For methods of observing dark matter directly in the laboratory, see \cite{oikonomouvergados}. 

In reference to late time acceleration, the cosmological models accounting for dark energy or late time acceleration largely take the two distinct ways. One is to add directly in the Einstein equations some non-standard matter component, with the most well known candidate being the cosmological constant $\Lambda$, and the corresponding model of cosmological evolution is known as the $\Lambda$-Cold-Dark-Matter($\Lambda$CDM). According to this model, the universe consists of ordinary matter ($\Omega_{m} \sim 4.9 \% $), cold dark matter ($\Omega_{DM} \sim 26.8 \% $), and dark energy ($\sim 68.3 \% $). Dark energy is the component of the matter energy tensor that generates the late time acceleration and one way to model this is provided by the $F(R)$ theories of gravity \cite{reviews1,reviews2,reviews3,reviews4,reviews5,reviews8,reviews9,
importantpapers1,importantpapers2,importantpapers3,importantpapers4,importantpapers5,
importantpapers6,importantpapers8,importantpapers9,importantpapers10,importantpapers11,importantpapers11a,
importantpapers12,importantpapers13,importantpapers14,importantpapers15,importantpapers17,
importantpapers18,importantpapers19,importantpapers20,sergeinojirimodel}, in which the right hand of the Einstein equations is directly modified due to the existence of terms coming from the geometry of the theory itself. For a list of reviews and important papers on this vast topic, see \cite{reviews1,reviews2,reviews3,reviews4,reviews5,reviews8,reviews9,
importantpapers1,importantpapers2,importantpapers3,importantpapers4,importantpapers5,
importantpapers6,importantpapers8,importantpapers9,importantpapers10,importantpapers11,importantpapers11a,
importantpapers12,importantpapers13,importantpapers14,importantpapers15,importantpapers17,
importantpapers18,importantpapers19,importantpapers20,sergeinojirimodel} and references therein. The challenge in modern cosmology is to find a unique theoretical framework to explain both late time and inflation at once. One such description was provided at first time by the Nojiri-Odintsov model  in \cite{sergeinojirimodel}, in which dark energy and also inflation was generated from a single model. For theories that can also consistently address the dark energy problem, see \cite{capo,capo1,peebles,faraonquin,tsujiintjd} and relevant references therein and for a recent study on unified dark energy with quintessential inflation \cite{quintense}, see \cite{saridakismyrzakulov}.

In a recent study \cite{oikonomoukaragiannakis}, we investigated the effect of collisional matter \cite{kleidis} on the late time evolution of the universe, in the context of $F(R)$ modified theories of gravities. The resulting picture of our investigation was that the behavior of the collisional matter filled universe, varied in a model dependent way, giving in some cases better fit to the $\Lambda$CDM model with respect to the non-collisional matter filled universe, always in the context of $F(R)$ gravities. Motivated by this rather vague result, in this paper we further scrutinize the effect of collisional matter in the cosmological evolution of $F(R)$ theories. Our approach is different with regards to our previous work, in which we were interested in the late time epoch. In the present work we shall consider the matter domination eras and also study the oscillatory behavior of dark energy. With regards to the matter domination era, we shall also examine the matter perturbations of the collisional universe and compare it to the non-collisional matter filled universe.

This paper is organized as follows: In section 1 we briefly review the fundamental features of $F(R)$ gravity, in section 2 we present necessary information for the exponential $F(R)$ gravity model which we shall use and also for the we give a brief account on the collisional matter essentials. Then, by introducing the necessary for our study variables, we explicitly re-calculate the corresponding equations of motion and we solve these numerically. We investigate the behavior of the dark energy density, the dark energy equation of state parameter, the effective equation of state parameter and also the Hubble parameter. We compare the results coming from a collisional matter filled plus radiation universe with those coming from a universe filled with ordinary matter plus radiation. We give special attention to dark energy oscillations in the process and investigate how the oscillations behave under the change of the parameters that characterize the collisional matter. In the end of section 2, we study in detail the behavior of matter perturbations and focus our investigation on the study of the growth factor as a function of the redshift $z$. In section 3, we discuss in detail the results of this article and combine these to our previous work results \cite{oikonomoukaragiannakis}. The conclusions follow in the end of the paper.

\section{Overview of Geometric Dark Energy and $F(R)$ Gravity Dynamics}

In order to maintain the article self contained, it worths recalling the essential features of the Jordan frame formulated $F(R)$ modified theories of gravity, considered in the metric formalism. For detailed work on these issues, see \cite{reviews1,reviews2,reviews3,reviews4,reviews5,importantpapers1,
importantpapers2,importantpapers3,importantpapers4,importantpapers5,importantpapers6,importantpapers8,importantpapers9,importantpapers10,importantpapers11,importantpapers11a,importantpapers12,importantpapers13} and references therein. It is assumed that the geometric background is that of a pseudo-Riemannian manifold, which locally is a Lorentz metric, which in our case is a flat Friedmann-Robertson-Walker metric (FRW hereafter), of the following form,
\begin{equation}\label{metricformfrwhjkh}
\mathrm{d}s^2=-\mathrm{d}t^2+a^2(t)\sum_i\mathrm{d}x_i^2
\end{equation}
The corresponding to this metric Ricci scalar is,
\begin{equation}\label{ricciscal}
R=6(2H^2+\dot{H}),
\end{equation}
where $H(t)$ denotes the Hubble parameter and the ``dot'' denotes differentiation with respect to the cosmological time $t$. In addition, it is assumed that the connection on this manifold is a torsion-less, symmetric, and
metric compatible affine connection, the Levi-Civita connection. The general action that describes $F(R)$ modified theories of gravity in the four dimensional Jordan frame is,
\begin{equation}\label{action}
\mathcal{S}=\frac{1}{2\kappa^2}\int \mathrm{d}^4x\sqrt{-g}F(R)+S_m(g_{\mu \nu},\Psi_m),
\end{equation}
with $\kappa^2=8\pi G$ and in addition $S_m$ contains all the matter content of the theory. Variation of the action (\ref{action}) with respect to the metric $g_{\mu \nu}$ results in the following equations of motion,
\begin{align}\label{modifiedeinsteineqns}
R_{\mu \nu}-\frac{1}{2}Rg_{\mu \nu}=\frac{\kappa^2}{F'(R)}\Big{(}T_{\mu
\nu}+\frac{1}{\kappa^2}\Big{(}\frac{F(R)-RF'(R)}{2}g_{\mu \nu}+\nabla_{\mu}\nabla_{\nu}F'(R)-g_{\mu
\nu}\square F'(R)\Big{)}\Big{)}.
\end{align}
where as usual, $F'(R)=\partial F(R)/\partial R$ and $T_{\mu \nu}$ denotes the energy momentum tensor corresponding to ordinary matter fields. As we already mentioned in the introduction, in the context of $F(R)$ theories, dark energy finds an elegant and self consistent description and it is named geometric dark energy. The reason for this can directly be extracted from the equations of motion above (\ref{modifiedeinsteineqns}). As it can be seen, the left hand side of these equations is identical to the Einstein equations of general relativity, while the right hand side is modified in a way that describes ordinary matter plus a contribution coming from a perfect fluid contribution, with purely geometric origin. Indeed we can write the equations of motion in the following form (for a very informative work on this see \cite{importantpapers13}),
\begin{equation}\label{geometridde}
R_{\mu \nu}-\frac{1}{2}Rg_{\mu \nu}=T_{\mu \nu}^{m}+T_{\mu \nu}^{curv}
\end{equation} 
with $T_{\mu \nu}^{m}$ being equal to,
\begin{equation}\label{tmnmat}
T_{\mu \nu}^{m}=\frac{1}{\kappa}\frac{T_{\mu \nu}}{F'(R)}
\end{equation}
and originating from ordinary matter perfect fluids, while $T_{\mu \nu}^{curv}$ is equal to,
\begin{equation}\label{tmnmatcurv}
T_{\mu \nu}^{curv}=\frac{1}{\kappa}\Big{(}\frac{F(R)-RF'(R)}{2}g_{\mu
\nu}+F'(R)^{;\mu \nu}(g_{\alpha \mu}g_{\beta \nu }-g_{\alpha \beta}g_{\mu \nu })\Big{)}
\end{equation}
The energy momentum tensor (\ref{tmnmatcurv}) is attributed to a perfect fluid of geometric origin, which describes the dark energy and is also called curvature fluid.

\section{Cosmological Evolution with Collisional Matter in Exponential Gravity}

\subsection{Viable Exponential Gravity with Curvature Corrections}

In principle an $F(R)$ gravity model has to pass a series of serious tests in order it can be considered a viable model. The constraints put by these tests are coming both from local astrophysical data and also from global, at a cosmological level, considerations, see for example \cite{reviews1,reviews2,reviews3,reviews4,reviews5,importantpapers1,
importantpapers2,importantpapers3,importantpapers4,importantpapers5,importantpapers6,importantpapers8,importantpapers9,importantpapers10,importantpapers11,importantpapers11a,importantpapers12,importantpapers13} and related references therein. In this paper we shall consider a viable exponential model, firstly studied in \cite{importantpapers11} by Cognola, Elizalde et al., which passes all the viability tests and also provides a very elegant theoretical framework for explaining late time acceleration and also inflation, at the early stages of the universe's evolution, mimicking the $\Lambda$CDM model at large curvature values. For studies on exponential models, the reader is referred to \cite{importantpapers11,importantpapers11a,exp1,exp2,exp3,exp4}. The model is described by the following $F(R)$ function,
\begin{equation}\label{expmodnocurv}
F(R)=R-2\Lambda \left ( 1-e^{\frac{R}{b\Lambda}} \right )
\end{equation}
with $\Lambda$ being the cosmological constant corresponding to present time, which is equal to $\Lambda \simeq 11.89 $eV$^2$ and $b$ is a free positive parameter which is assumed to be $b\simeq 1$ (we adopt the numerical values of reference \cite{exp1}). Regardless the good viability properties of the model (\ref{expmodnocurv}), there are some complications that arise in the process of the cosmological evolution and particularly during the matter domination era. Particularly, the higher derivatives of the Hubble parameter diverge, an issue that originates from the dark energy oscillations during the matter phase \cite{exp0,exp1,exp3}. In fact for large values of the redshift, there occur high  frequency dark energy oscillations, and also the derivatives of the dark energy density, denoted as $\rho_{DE}$, take large values, which leave this imprint in the dark energy equation of state parameter $\omega_{DE}$. Actually, the dark energy oscillations become more pronounced (higher frequency) in $F(R)$ modified gravity models that mimic more the $\Lambda$CDM model, which practically speaking occurs when $F''(R)$ is close to zero. Thus our fundamental function of the exponential model (\ref{expmodnocurv}) to correct Einstein's gravity in the small curvature regime, causes undesirable features in the high curvature regime. A way out of this is offered by directly modifying the $F(R)$ gravity Lagrangian, in order to stabilize dark energy oscillations, in such a way so that the frequency of the oscillations becomes constant \cite{exp1,exp3}. A consistent modification was proposed in \cite{exp1}, so that the curvature corrected $F(R)$ Lagrangian reads,
\begin{equation}\label{expmodnocurvcorr}
F(R)=R-2\Lambda \left ( 1-e^{\frac{R}{b\Lambda}} \right )-\tilde{\gamma}\Lambda \left (\frac{R}{3\tilde{m}^2} \right )^{1/3}
\end{equation}
with $\Lambda=7.93\tilde{m}^2$ and $\tilde{\gamma}=1/1000$ \cite{exp1}. The addition of the curvature correction stabilizes the dark energy oscillations while keeping all the viability features of the model intact. For example the Minkowski spacetime still is the flat space solution and the effects of the curvature correction vanish in the de-Sitter epoch, provided $\tilde{\gamma}\ll (\tilde{m}^2/\Lambda)^{1/3} $, which obviously is satisfied for the adopted value of $\tilde{\gamma}$.

\subsection{Dark Energy and Equation of State Oscillations with Collisional Matter Fluids-Cosmological Evolution}

The detailed description of a perfect matter fluid with interactions was given in \cite{kleidis,fock} and reference therein (see also \cite{oikonomoukaragiannakis}). It worths recalling in brief, the basic features which we shall use in this article. The basic assumption to be made is that the collisional matter has a total mass-energy density, denoted as $\varepsilon_m$, which receives two contributions in it's functional form and reads,
\begin{equation}\label{generenergydensit}
 \varepsilon_m=\rho_m+\rho_m\Pi
\end{equation}
In the expression above, $\rho_m$ stands for the rest mass density which remains unaffected by the internal motions of the gravitational cosmic
fluid and the part containing the potential energy $\Pi$, is what renders the matter fluid collisional \cite{kleidis,fock}. The collisional matter is considered to be a prefect fluid with equation of state, 
\begin{equation}\label{eqnstate}
 p_m=w\rho_m
 \end{equation}
with the equation of state parameter taking values $0< w < 1$. The potential energy density is assumed to be of the form \cite{kleidis,fock},
\begin{equation}\label{potenerg}
\Pi=\Pi_0+w\ln (\frac{\rho_m}{\rho_m^{(0)}})
\end{equation}
with $\rho_m^{(0)}$ and $\Pi_0$ denoting the present values of the motion invariant mass energy density and of 
the potential energy, respectively. Hence, the total energy density of the gravitational fluid
is equal to,
\begin{equation}\label{totenergydens}
\varepsilon_m=\rho_m\Big{[}1+\Pi_0+w \ln (\frac{\rho_m}{\rho_m^{(0)}})\Big{]}
\end{equation}
In addition and due to the continuity equation for the collisional matter fluid (see the book of Fock \cite{fock}, pages 90-93), the continuity equation of the gravitational fluid in a spatially flat FRW metric takes the following form,
\begin{equation}\label{contquenne}
\dot{\varepsilon}_m+3\frac{\dot{a}}{a}(\varepsilon_m+p_m)=0
\end{equation}
which leads to \cite{kleidis},
\begin{equation}\label{evlotuniomatt}
\rho_m=\rho_{m}^{(0)}\big{(}\frac{a_0}{a}\big{)}^3
\end{equation}
with $a$ and $a_0$ denoting the scale factor and it's present value respectively. Finally the value of $\Pi_0$ is equal to \cite{kleidis},
\begin{equation}\label{pio}
\Pi_0=\Big{(}\frac{1}{\Omega_M}-1\Big{)}
\end{equation}
which we shall use in our numerical calculations. In the next section we shall go through the equations of cosmological evolution and see in detail how these are modified in the presence of a collisional matter fluid with total mass-energy density given by relation (\ref{totenergydens}).

\subsection{Modification of Standard Cosmological Evolution in the Presence of Collisional Matter Fluid}

Our basic intention and core subject of this section, is to study the cosmological evolution of a universe filled with collisional matter and radiation, with special emphasis in the redshift region which corresponds to matter domination and later. This is in order to study the dark energy oscillations during this era, which as we already mentioned in the previous sections, can have a significant value during this era and at low redshifts in general. We shall express the quantities entering the FRW equations in terms of the redshift parameter $z=1/a-1$ and also use new variables, appropriate for the study of dark energy oscillations. We shall perform a numerical analysis of the modified FRW equations for the type of collisional matter described in the previous section, and thoroughly study the behavior of the dark energy density $\Omega_{DE}(z)$, the modified Hubble parameter, $H(z)$ and of the effective equation of state parameter $\omega_{eff}$, which we re-express later in terms of the new variables we shall introduce. For the numerical analysis, we shall take into account values of the redshift $z\leq 10$, which covers the last stages of the matter domination era, which started at approximately $z\sim 3000$ and is considered to end at $z\simeq 3$ (this can vary from model to model). After that the late time acceleration era starts. The motivation for us to use the redshift as our central parameter of our analysis comes from the fact that, the cosmological distances in standard cosmology are determined by using standard candles as references, such as TypeIa supernovae or Gamma Ray Bursts (GRB hereafter wherever used). The TypeIa supernovae correspond to a redshift with values between $0<z<1.7$. In addition, GRBs are visible in much more higher redshifts, up to $z=6$, which corresponds to the late stages of the matter domination era. This is why we shall use values of the redshift with $0\leq z\leq 10$, in order to take into account future observational data coming from GRBs corresponding to high redshifts with $z\geq 6$.  

In order to proceed in the description of the cosmological evolution, we rewrite the cosmological equations in a specific form. Particularly, the FRW equations of motion (\ref{modifiedeinsteineqns}) can be written in the following form,
\begin{align}\label{eq:flrw}
& 3F'H^2=k^2\rho_{matt}+\frac{1}{2}(F'R-F)-3H\dot{F'}\\ \notag &
-2F'\dot{H}=k^2\left (\rho_{matt}+P_{matt} \right )+\ddot{F}-H\dot{F}
\end{align}
with $\rho_{matt}$ being the total mass-energy density containing all the matter fluids. In the present case the matter fluids we shall take into account consist of collisional matter and relativistic matter (radiation), so $\rho_{matt}$ is in our case,
 \begin{equation}\label{totalmattenergdensmf}
\rho_{matt}=\varepsilon_m+\rho_{r}^{(0)}a^{-4}
\end{equation}
and by taking into account relations (\ref{totenergydens}) and (\ref{evlotuniomatt}), equation (\ref{totalmattenergdensmf}) becomes,
\begin{equation}\label{totalmattenergdensmf1}
\rho_{matt}=\rho_m^{(0)}a^{-3}\Big{[} 1+\Pi_0+3w \ln (a) \Big{ ]}+\rho_{r}^{(0)}a^{-4}
\end{equation}
In addition, $P_{matt}$ is the corresponding total pressure corresponding to all the matter fluids. The first equation of the FRW equations in relation (\ref{eq:flrw}) can be written as, 
\begin{equation}\label{eq:modifiedeinsteineqns2}
  H^2-(F'-1)\left (H\frac{\mathrm{d}H}{\mathrm{d}\ln{a}}+H^2 \right )+\frac{1}{6}(F-R)+H^2F''\frac{\mathrm{d}R}{\mathrm{d}\ln{a}}=\frac{\rho_{matt}}{3},
\end{equation}
where $R$ is the Ricci scalar which for the purposes of this paper, can be written as,
\begin{equation}\label{eq:ricciscal2}
  R=12H^2+6H\frac{\mathrm{d}H}{\mathrm{d}\ln{a}},
\end{equation}
In order to provide general formulas that hold true for any type of collisional matter, we shall assume that the total matter-energy density takes the following general form,
\begin{equation}
  \rho_{matt}=\rho_m^{(0)}g(a)+\rho_{r}^{(0)}a^{-4}=\rho_m^{(0)}\left( g(a)+\chi a^{-4}\right ),
\end{equation}
with $\rho_m^{(0)}$ and $\rho_r^{(0)}=\chi\rho_m^{(0)}$ being the present values of the mass-energy densities of matter and radiation respectively, and $\chi\simeq3.1\times10^{-4}$ being defined as the ratio $\rho_r^{(0)}/\rho_m^{(0)}$. So practically, the collisional nature of the non-relativistic matter is described by the function $g(a)$, which in the case of collisional matter described in the previous section, this is equal to,
\begin{equation}\label{fgd}
g(a)=a^{-3}\Big{[} 1+\Pi_0-3w \ln (a) \Big{ ]}
\end{equation}
In order to consistently describe the dark energy oscillations, we re-express the cosmological equations (\ref{eq:modifiedeinsteineqns2}) in terms of new parameters which have vanishing values at the high redshift limits, where the $F(R)$ modifications are negligible \cite{importantpapers4},
\begin{subequations}
  \begin{align}
    y_H&\equiv\frac{\rho_{DE}}{\rho_m^{(0)}}\frac{H^2}{\tilde{m}^2}-g(a)-\chi a^{-4}\label{eq:yH}\\
    y_R&\equiv\frac{R}{\tilde{m}^2}-\frac{\mathrm{d}g(a)}{\mathrm{d}\ln{a}}\label{eq:yR}
  \end{align}
\end{subequations}
with $\rho_{DE}$ the energy density of dark energy. So the new quantity which quantifies the cosmological evolution is, $y_H(z)$, which is the scaled dark energy density, and scaled by the factor $\rho_m^{(0)}$. Dividing the equations (\ref{eq:modifiedeinsteineqns2}) by $\tilde{m}^2$, and upon using, 
\begin{equation}
  \frac{H^2}{\tilde{m}^2}-\frac{\rho_M}{2\tilde{m}^2}=\frac{H^2}{\tilde{m}^2}-g(a)-\chi a^{-4}=y_H
\end{equation}
from (\ref{eq:yH}), we solve the resulting expression with respect to $\frac{1}{\tilde{m}^2}\frac{dR}{d\ln{a}}$. The final result is,
\begin{equation}\label{eq:dRdlna}
  \frac{1}{\tilde{m}^2}\frac{\mathrm{d}R}{\mathrm{d}\ln{a}}=\left[-y_H+(F'-1)(\frac{H}{\tilde{m}^2}\frac{\mathrm{d}H}{\mathrm{d}\ln{a}}+
    \frac{H^2}{\tilde{m}^2})-\frac{1}{6\tilde{m}^2}(F-R)\right]\frac{1}{H^2F''}
\end{equation}
Upon differentiation of relation (\ref{eq:yR}), with respect to $\ln{a}$ we get,
\begin{equation}
  \frac{\mathrm{d}R}{\mathrm{d}\ln{a}}=\frac{1}{\tilde{m}^2}\frac{\mathrm{d}R}{\mathrm{d}\ln{a}}-\frac{\mathrm{d}^2g(a)}{\mathrm{d}\ln{a}^2},
\end{equation}
in which we substitute (\ref{eq:dRdlna}),
\begin{equation}\label{eq:dyR}
  \frac{\mathrm{d}y_R}{\mathrm{d}\ln{a}}=-\frac{\mathrm{d}^2g(a)}{\mathrm{d}\ln{a}^2}+\left[-y_H+(F'-1)(\frac{H}{\tilde{m}^2}\frac{\mathrm{d}H}{\mathrm{d}\ln{a}}
    +\frac{H^2}{\tilde{m}^2})-\frac{1}{6\tilde{m}^2}(F-R)\right]\frac{1}{H^2F''}
\end{equation}
Upon differentiation of (\ref{eq:yH}) with respect to $\ln{a}$, we get,
\begin{eqnarray}\label{eq:dHlna}
  \frac{\mathrm{d}y_H}{\mathrm{d}\ln{a}}&=&2\frac{H}{\tilde{m}^2}\frac{\mathrm{d}H}{\mathrm{d}\ln{a}}-\frac{\mathrm{d}g(a)}{\mathrm{d}\ln{a}}+4\chi
    a^{-4}\nonumber\\
  \Rightarrow\frac{H}{\tilde{m}^2}\frac{\mathrm{d}H}{\mathrm{d}\ln{a}}&=&\frac{1}{2}\frac{\mathrm{d}y_H}{\mathrm{d}\ln{a}}+\frac{1}{2}
    \frac{\mathrm{d}g(a)}{\mathrm{d}\ln{a}}-2\chi a^{-4}\nonumber\\
  \Rightarrow\frac{H}{\tilde{m}^2}\frac{\mathrm{d}H}{\mathrm{d}\ln{a}}+\frac{H^2}{\tilde{m}^2}&=&\frac{1}{2}\frac{\mathrm{d}y_H}{\mathrm{d}\ln{a}}
    +\frac{1}{2}\frac{\mathrm{d}g(a)}{\mathrm{d}\ln{a}}+y_H+g(a)-\chi a^{-4},
\end{eqnarray}
where we made use of relation (\ref{eq:yH}). Combining equations (\ref{eq:dHlna}) and (\ref{eq:dyR}), we get the following expression, 
\begin{eqnarray}\label{eq:dyR2}
  \frac{\mathrm{d}y_R}{\mathrm{d}\ln{a}}&=&-\frac{\mathrm{d}^2g(a)}{\mathrm{d}\ln{a}^2}+
    \left[-y_H+(F'-1)\left(\frac{1}{2}\frac{\mathrm{d}y_H}{\mathrm{d}\ln{a}}+\frac{1}{2}\frac{\mathrm{d}g(a)}{\mathrm{d}\ln{a}}
    +y_H+g(a)-\chi a^{-4}\right)\right.\nonumber\\
    &&\left.-\frac{1}{6\tilde{m}^2}(F-R)\right]\frac{1}{\tilde{m}^2F''(y_H+g(a)+\chi a^{-4})}.
\end{eqnarray}
Differentiating relation (\ref{eq:dRdlna}) with respect to $\ln{a}$, results to,
\begin{equation}\label{eq:dyH} 
  \frac{\mathrm{d}y_H}{\ln{a}}=\frac{2H}{\tilde{m}^2}\frac{\mathrm{d}H}{\mathrm{d}\ln{a}}-\frac{\mathrm{d}g(a)}{\mathrm{d}\ln{a}}+4\chi a^{-4}.
\end{equation}
From (\ref{eq:ricciscal2}) and (\ref{eq:yH}) we get,
\begin{equation}
  \frac{2H}{\tilde{m}^2}\frac{\mathrm{d}H}{\ln{a}}=\frac{R}{3\tilde{m}^2}-\frac{4H^2}{\tilde{m}^2}=\frac{R}{3\tilde{m}^2}-4y_H-4g(a)-4\chi
a^{-4},
\end{equation}
so relation (\ref{eq:dyH}) becomes,
\begin{eqnarray}\label{eq:dyH2}
  \frac{\mathrm{d}y_H}{\mathrm{d}\ln{a}}&=&\frac{R}{3\tilde{m}^2}-\frac{\mathrm{d}g(a)}{\mathrm{d}\ln{a}}-4y_H-4g(a)\nonumber\\
    &=&\frac{y_R}{3}-4y_H-\frac{2\mathrm{d}g(a)}{3\mathrm{d}\ln{a}}-4g(a)
\end{eqnarray}
and in addition the Ricci scalar can be written as,
\begin{equation}
  R=3\tilde{m}^2\left(4y_H+4g(a)+\frac{\mathrm{d}y_H}{\mathrm{d}\ln{a}}+\frac{\mathrm{d}g(a)}{\mathrm{d}\ln{a}}\right)
\end{equation}
Upon differentiation of relation (\ref{eq:dyH2}) with respect to $\ln{a}$,
\begin{equation*}
  \frac{\mathrm{d}^2y_H}{\mathrm{d}\ln{a}^2}=\frac{\mathrm{d}y_R}{3\mathrm{d}\ln{a}}-\frac{4\mathrm{d}y_H}{\mathrm{d}\ln{a}}
    -\frac{2}{3}\frac{\mathrm{d}^2g(a)}{\mathrm{d}\ln{a}^2}-4\frac{\mathrm{d}g(a)}{\mathrm{d}\ln{a}}
\end{equation*}
and by making use of relation (\ref{eq:dyR2}), we obtain,
\begin{eqnarray}\label{eq:FRform}
  &&\frac{\mathrm{d}^2y_H}{\mathrm{d}\ln{a}^2}+\left(4+\frac{1-F'}{6\tilde{m}^2F''(y_H+g(a)+
    \chi a^{-4})}\right)\frac{\mathrm{d}y_H}{\mathrm{d}\ln{a}}
    +\left(\frac{2-F'}{3\tilde{m}^2F''(y_H+g(a)+\chi a^{-4})}\right)y_H
  \nonumber\\
  &&+\left(\frac{\mathrm{d}^2g(a)}{\mathrm{d}\ln{a}^2}+4\frac{\mathrm{d}g(a)}{\mathrm{d}\ln{a}}
    +\frac{(1-F')\left(3\frac{\mathrm{d}g(a)}{\mathrm{d}\ln{a}}+6g(a)-6\chi a^{-4}\right)+\frac{F-R}{\tilde{m}^2}}
      {18\tilde{m}^2F'' (y_H+g(a)+\chi a^{-4})}\right)=0
\end{eqnarray}
Finally we express all quantities as functions of the redshift $z$, and order to do that, we shall make use of the following expressions,
\begin{subequations}
  \begin{align}
    \frac{\mathrm{d}}{\mathrm{d}\ln{a}}&=-(z+1)\frac{\mathrm{d}}{\mathrm{d}z},\\
    \frac{\mathrm{d}^2}{\mathrm{d}\ln{a}^2}&=(z+1)\frac{\mathrm{d}}{\mathrm{d}z}+(z+1)^2\frac{\mathrm{d}^2}{\mathrm{d}z^2}.
  \end{align}
\end{subequations}
Then, by applying the above formulas in relation (\ref{eq:FRform}), we obtain the following final form of the differential equation that describes the cosmological evolution of the universe filled with radiation and collisional matter,
\begin{eqnarray}
  &&\frac{\mathrm{d}^2y_H}{\mathrm{d}z^2}+
    \frac{1}{(z+1)}\left(-3-\frac{F'(R)}{6\tilde{m}^2F''(R)(y_H+g(z)+\chi(z+1)^4)}\right)\frac{\mathrm{d}y_H}{\mathrm{d}z}
  \nonumber\\
  &&+\frac{1}{(z+1)^2}\frac{1-F'(R)}{3\tilde{m}^2F''(R)(y_H+g(Z)+\chi(z+1)^4)}y_H+\left(\frac{
    \mathrm{d}^2g(z)}{\mathrm{d}z^2}-\frac{3}{(z+1)}\frac{\mathrm{d}g(z)}{\mathrm{d}z}\right.
  \nonumber\\
  &&\left.-\frac{1}{(z+1)}\frac{F'(R)\left(-(z+1)\frac{\mathrm{d}g(z)}{
    \mathrm{d}z}+2g(z)-2\chi(z+1)^4\right)+\frac{F}{3\tilde{m}^2}}{6\tilde{m}^2F''(R)(y_H-g(z)+\chi(z+1)^4)}\right)=0
\end{eqnarray}
which is a second order differential equation in terms of the scaled dark energy $y_H$. In the rest of this section we solve numerically this equation for the matter profile $g(a)$ given in relation (\ref{fgd}) and compare the dark energy oscillations and the cosmological evolutions of collisional matter in comparison to the non-collisional one. It worths to present briefly the set of initial conditions and also the values of the various parameters we shall use. Specifically, the present value of the matter density is $\Omega_M=0.279$, the equation of state parameter of collisional matter shall be assumed to have two different values, $w=0.1$ and $w=0.7$, and the initial conditions are taken to be identical to those used in reference \cite{exp1},
\begin{equation}\label{initialcond}
y_H(z)\mid_{z=z_{f}}=\frac{\Lambda }{3\tilde{m}^2}\left(1+\frac{z_{f}+1}{1000}\right),{\,}{\,}{\,}y_{H}'(z)\mid_{z=z_{f}}=\frac{\Lambda }{3\tilde{m}^2}\frac{1}{1000}
\end{equation}
with $z_{f}=10$ and $\Lambda $, $\tilde{m}^2$ being defined in the previous sections. 

\begin{figure}[h]
\centering
\includegraphics[width=15pc]{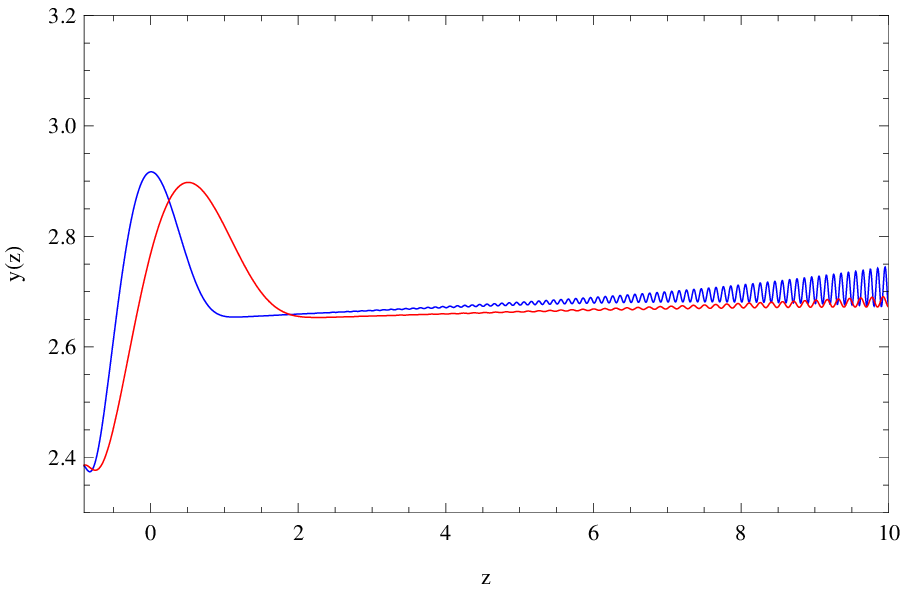}
\includegraphics[width=15pc]{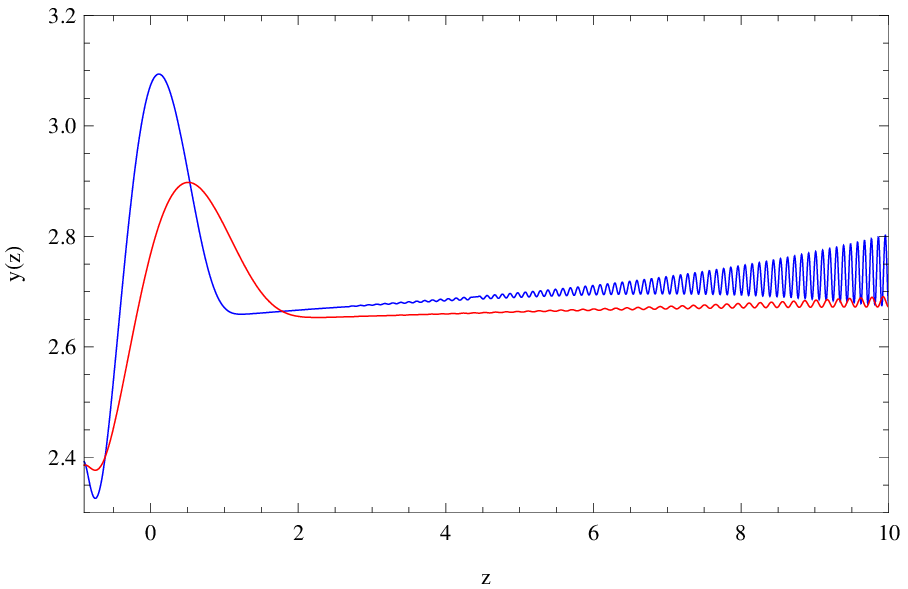}
\caption{Comparison of the scaled dark energy density $y_H(z)=\frac{\rho_{DE}}{\rho_m^{(0)}}$ over z, for $w=0.1$ (left) and $w=0.7$ (right). The red line corresponds to non-collisional matter while the blue corresponds to collisional matter}\label{ycomp}
\end{figure}

\begin{figure}[h]
\centering
\includegraphics[width=15pc]{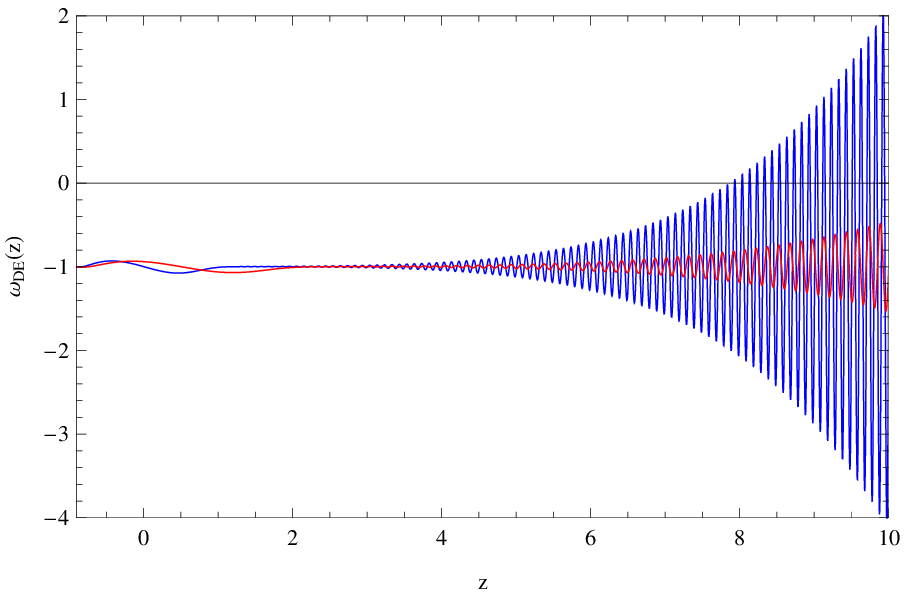}
\includegraphics[width=15pc]{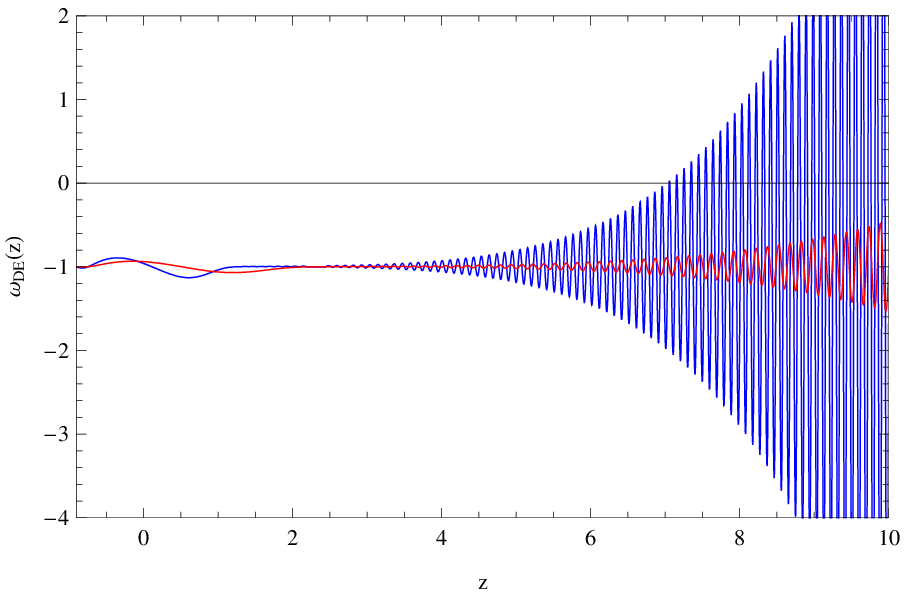}
\caption{Comparison of the dark energy equation of state parameter $\omega_{DE}(z)$ over z, for $w=0.1$ (left) and $w=0.7$ (right). The red line corresponds to non-collisional matter while the blue corresponds to collisional matter}\label{omegadecomp}
\end{figure}

We start off with the comparison of the function $y_H(z)$ for collisional and non-collisional matter and in Fig. \ref{ycomp} we present of the numerical analysis we performed, for $w=0.1$ (left) and $w=0.7$ (right) with the blue line corresponding to collisional matter with $g(a)$ given in (\ref{fgd}) and the red line corresponding to non-collisional matter ($g(a)=a^{-3}$).
As we can see, the behavior of the scaled dark energy density $y_H(z)=\frac{\rho_{DE}}{\rho_m^{(0)}}$ for collisional matter is more oscillatory in reference to non-collisional matter, and this phenomenon gets more pronounced as $w$ increases. The latter is rather normal, because the more $w$ approaches the value zero, the collisional nature of matter disappears. In reference to dark energy oscillations, we shall compare the dark energy equation of state parameter $\omega_{DE}=P_{DE}/\rho_{DE}$ for collisional and non-collisional matter for various $w$ values. It worths providing the formula from which the dark energy density is given, which is the following,
\begin{equation}\label{deeqnstateprm}
\omega_{DE}(z)=-1+\frac{1}{3}(z+1)\frac{1}{y_H(z)}\frac{\mathrm{d}y_H(z)}{\mathrm{d}z}
\end{equation}
In Fig. \ref{omegadecomp} we present the comparison plots for the behavior of the dark energy equation of state parameter for collisional matter (blue) and for non-collisional matter (red) with $w=0.1$ (left) and $w=0.7$ (right). 
As we can see, the oscillatory behavior of dark energy is highly pronounced as $w$ increases and also it is more oscillatory in comparison to non-collisional matter.

\begin{figure}[h]
\centering
\includegraphics[width=15pc]{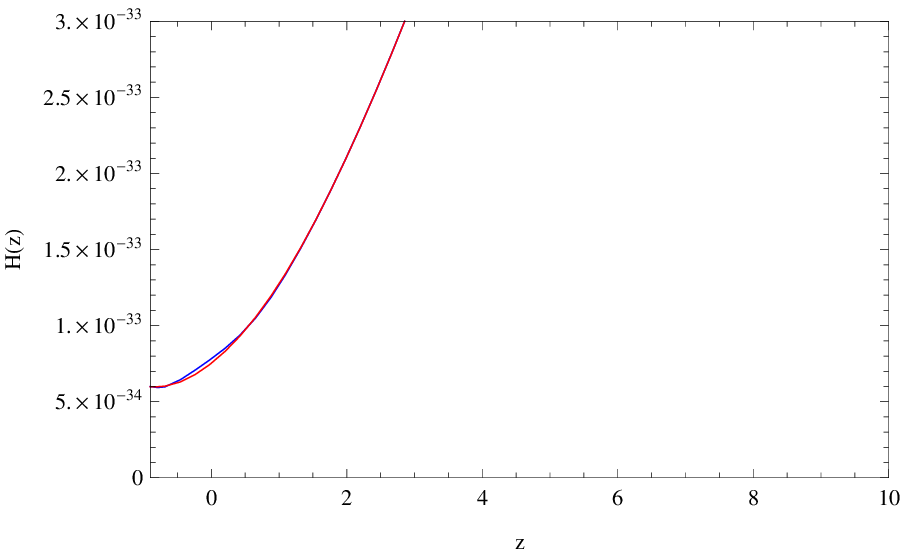}
\includegraphics[width=15pc]{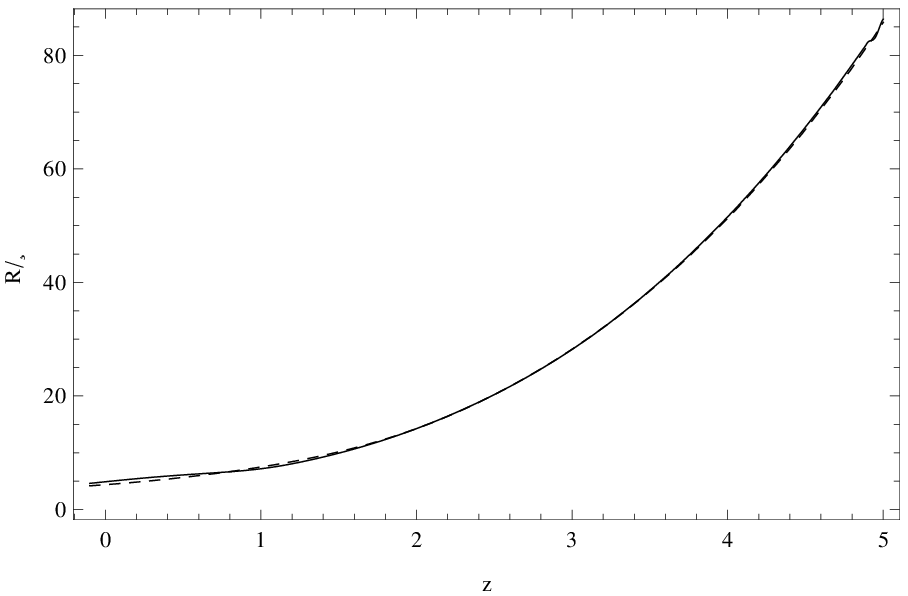}
\caption{Comparison of the Hubble parameter $H(z)$ over z (left) and of the Ricci scalar $R(z)$ (right), for $w=0.7$. The red line corresponds to non-collisional matter while the blue corresponds to collisional matter}\label{hubcurv}
\end{figure}

\begin{figure}[h]
\centering
\includegraphics[width=15pc]{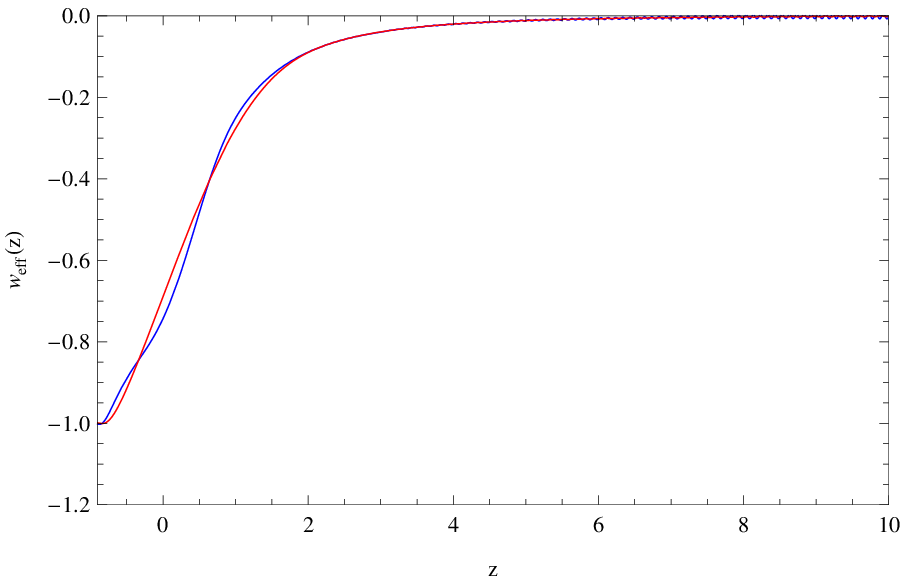}
\includegraphics[width=15pc]{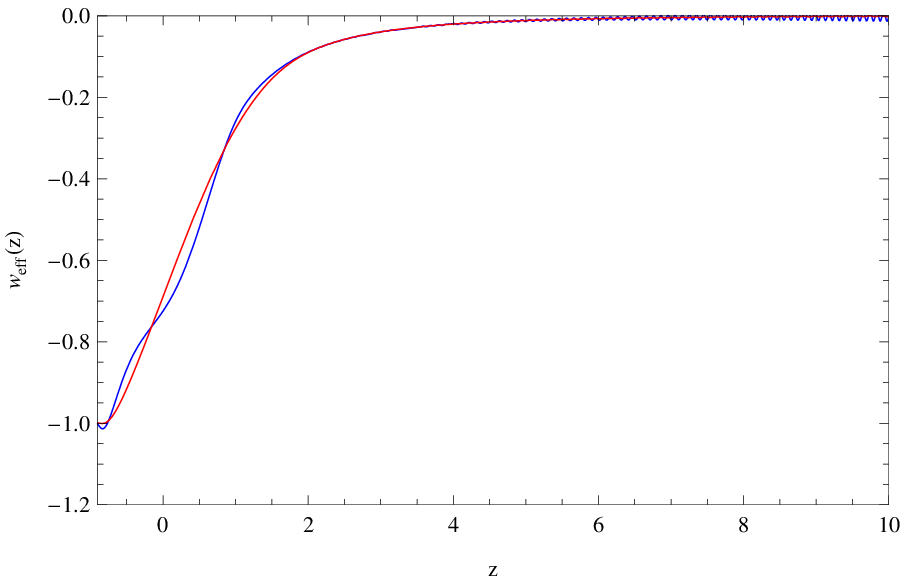}
\caption{Comparison of the effective equation of state parameter $\omega_{eff}(z)$ over z, for $w=0.1$ (left) and $w=0.7$ (right). The red line corresponds to non-collisional matter while the blue corresponds to collisional matter}\label{weffcomp}
\end{figure}

Let us now turn our focus to the cosmological evolution of the universe filled with matter and radiation and study the parameters involved in the cosmological evolution, namely the Hubble parameter $H(z)$, the scalar curvature $R(z)$ and the effective (total) equation of state corresponding to all the perfect fluids of the universe, namely radiation, matter (collisional or not) and dark energy. We start off with the Hubble parameter, which can be expressed as a function of the scaled dark energy density $y_H(z)$. Indeed, using equation (\ref{eq:yH}), the Hubble parameter is,
\begin{equation}\label{hubblepar}
H(z)=\sqrt{\tilde{m}^2y_H(z)+g(a(z))+\chi (z+1)^{4}}
\end{equation} 
and by using the numerical solution for $y_H(z)$ in the left plot of Fig. \ref{hubcurv} we present the evolution of the Hubble parameter as a function of $z$ for $w=0.7$. As we can see, the evolution of a universe filled with collisional matter is almost identical to the evolution of a universe filled with non-collisional matter. The same result can be reached if we compare the plots of the Ricci scalar as a function of $z$. The Ricci scalar can be expressed as a function of $y_H(z)$ as follows,
\begin{equation}
  R=3\tilde{m}^2\left(4y_H+4g(z)-(z+1)\frac{\mathrm{d}y_H}{\mathrm{d}z}-(z+1)\frac{\mathrm{d}g(z)}{\mathrm{d}z}\right).
\end{equation} 
and in the right plot of Fig. \ref{hubcurv} we compare it's evolution for collisional (blue) and non-collisional matter (red). 
Finally a similar behavior to the aforementioned two parameters appears if we consider the effective equation of state parameter. In Fig. (\ref{weffcomp}) we compare the effective equation of state parameters $\omega_{eff}=P_{tot}/\rho_{tot}$ for collisional (blue) and non-collisional (red) matter, for $w=0.1$ (left) and $w=0.7$ (right). The effective equation of state parameter is written in terms of the Hubble parameter $H(z)$ as follows,
\begin{equation}\label{effeqnofstateform}
\omega_{eff}(z)=-1+\frac{2(z+1)}{3H(z)}\frac{\mathrm{d}H(z)}{\mathrm{d}z}
\end{equation}
Notice that the total equation of state parameter never crosses the phantom divide line, while the dark energy equation of state parameter has strong oscillations around the phantom divide line, between the quintessence and phantom region. Recall that the value $w=-1$ \cite{reviews1},  corresponds to a cosmological constant perfect fluid equation of state, and the $-1<w<0$, corresponds to quintessence energy. Also the values $w<-1$ correspond to the phantom energy region.

It worths making some contact with some experimental values of the physical quantities we studied in this section. In Table 1 we present the present time ($z=0$) observational data for the dark energy density $\Omega_{DE}(z)$ and the dark energy equation of state parameter $\omega_{DE}(z)$ and also their obtained values for a universe filled with collisional and non-collisional matter fluids. 
 \begin{center} 
    \begin{tabular}{ | p{5cm} | p{3cm}  | p{3cm}  |}
    \hline
    Observational Quantities for $z=0$  & $\Omega_{DE}(z)$ & $\omega_{DE}(z)$  \\ \hline
    Collisional Matter Fluid for $w=0.01$ & $0.743501$ & $-1.00259$  
     \\ \hline
     Collisional Matter Fluid for $w=0.1$ & $0.744645$ & $-0.998188$ 
     \\ \hline
    Collisional Matter Fluid for $w=0.7$ & $0.754478$ & $-0.960615$ 
    \\ \hline 
    Non-Collisional Matter Fluid  & $0.734581$ & $-0.938451$  
    \\ \hline 
    Observational Values  & $0.721\pm 0.015$ & $-0.972\pm0.06$  
    \\ \hline 
    \end{tabular}
    \\
    \bigskip 
    Table 1: Values of $\Omega_{DE}(z)$ and $\omega_{DE}(z)$, corresponding to collisional and non-collisional matter fluids and also current observed values.
\end{center}
We have to notice two things, which we shall discuss further on in a later section. Firstly notice that there exist values of $w$ for which the dark energy equation of state parameter crosses the phantom divide line. Secondly, the $w=0.7$ collisional model gives better estimation with respect to $\omega_{DE}(z)$, but a bit worse for $\Omega_{DE}(z)$, in comparison to the non-collisional matter universe and always in reference to the observational data. We come to this point again in a later section in which we shall further discuss the cosmological implications of our numerical analysis results in detail. Before we close this section let us note that apart from the scaled dark energy density and the dark energy equation of state parameters, all the other parameters we examined generated almost indistinguishable cosmological evolution for a universe filled with collisional or non-collisional matter. Therefore, in order to have a consistent description for the evolution of the universe with various matter fluids we need to study other physical quantities that distinguish the evolution of different matter fluids. Such quantities can be provided by studying the matter density perturbations, which is the subject of the following section.

\subsection{Future Cosmological Evolution-Matter Density Perturbations}

\begin{figure}[h]
\centering
\includegraphics[width=15pc]{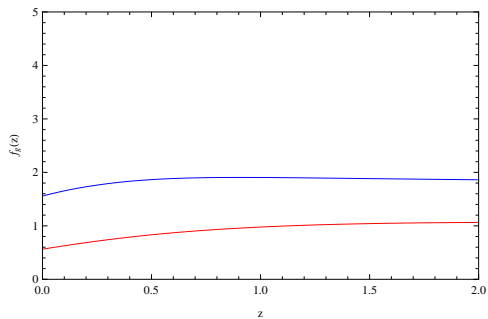}
\caption{Plots of the growth factor $f_g(z)=\frac{\mathrm{d}\ln \delta}{\mathrm{d}\ln a}$, which quantifies the matter perturbations during the matter era, as a function $z$, for $k=0.1$Mpc$^{-1}$ and $w=0.7$. The blue line corresponds to collisional matter and the red to non-collisional matter respectively.}\label{fig:oikonomouplots1}
\end{figure}

A consistent criterion to distinguish the cosmological evolution of different $F(R)$ gravities is the cosmological perturbation theory, in the context of which, even though two models may cause similar cosmological evolution, they can be distinguished. Practically this happens because cosmological perturbations actually differentiates the evolution of each model from the background evolution so this gives us information for this differentiation from the background evolution \cite{matsumoto}. In this paper we shall be interested in matter density perturbations, which are consistently calculated in the sub-horizon approximation, in the context of which the theory is rendered consistent with Newtonian gravity \cite{matsumoto}. The sub-horizon approximation qualitatively means that co-moving wavelengths $\lambda =a/k$ with the spacelike hypersurface that describes the evolution, are considered to be much shorter that the Hubble radius $H^{-1}$ corresponding to this hypersurface (notice that, loosely speaking, this condition is reminiscent to the geometric optics approximation in optics). This condition is quantitatively expressed by the following inequality \cite{matsumoto},
\begin{equation}\label{subhorapprx}
\frac{k^2}{a^2}\gg H^2
\end{equation}
where $k$ and $a$ are the wavenumber an the scale factor respectively. Note that the sub-horizon approximation breaks down during the radiation era and earlier in cosmic time. So we focus our interest on the matter era and onwards in cosmic time. In the case of collisional matter, the matter density perturbations are described by the parameter $\delta =\frac{\delta \varepsilon_m}{\varepsilon_m}$, with $\varepsilon_m$ the total mass-energy density of the collisional cosmic fluid under consideration (see relation (\ref{totenergydens})).  The matter energy perturbation $\delta$ satisfies the following equation \cite{exp1}, 
\begin{equation}\label{matterperturb}
\ddot{\delta}+2H\dot{\delta}-4\pi G_{eff}(a,k)\varepsilon_m\delta =0
\end{equation}
with $G_{eff}(a,k)$ standing for the effective gravitational constant of the modified gravity $F(R)$ theory given by \cite{exp1},
\begin{equation}\label{geff}
G_{eff}(a,k)=\frac{G}{F'(R)}\Big{[}1+\frac{\frac{k^2}{a^2}\frac{F''(R)}{F'(R)}}{1+3\frac{k^2}{a^2}\frac{F''(R)}{F'(R)}} \Big{]}
\end{equation}
with $G$ the gravitational constant of Newtonian gravity. We shall express every cosmic time dependent quantity appearing in the matter perturbations differential equation (\ref{matterperturb}) as a function of the redshift $z$ and also we re-express the differential equation to describe the quantity $f_g(z)=\frac{\mathrm{d}\ln \delta}{\mathrm{d}\ln a}$, which is called growth factor. By doing so, and also taking into account the following relations,
\begin{align}\label{basicrelations}
& \dot{H}=\frac{\mathrm{d}H}{\mathrm{d}z}(z+1)H(z), \\ \notag &
\dot{\delta}=H\dot{f_g}\delta \\ \notag &
\ddot{\delta }=\dot{H}f_g\delta+H\dot{f_g}\delta+Hf_g\dot{\delta},
\end{align}
the differential equation (\ref{matterperturb}), in the presence of collisional matter is written as follows,
\begin{equation}\label{presenceofcoll}
\frac{\mathrm{d}f_g(z)}{\mathrm{d}z}+\Big{(}\frac{1+z}{H(z)}\frac{\mathrm{d}H(z)}{\mathrm{d}z}-2-f_g(z)\Big{)}\frac{f_g(z)}{1+z}+\frac{4\pi}{G}\frac{G_{eff}(a(z),k)}{(z+1)H^2(z)}\varepsilon_m=0
\end{equation}
where we normalized the Newton's constant to the present one as $G_{eff}/G$ and $\varepsilon_m$ is given in relation (\ref{totenergydens}). In addition we can see that the collisional nature of matter modifies explicitly the matter perturbations, especially in the matter domination era. Notice that the presence of the term $G_{eff}(a(z),k)$ determines how the $F(R)$ gravity modifies the matter perturbations in an implicit way. Moreover, the effective gravitational constant $G_{eff}(a(z),k)$ has an explicit dependence on the wavenumber $k$, a feature that is absent in general relativity. This $k$-dependence should be taken into account during the computation of the CMB power spectrum, because the current spectrum is computed by using general relativity considerations \cite{exp1}. We shall numerically solve the differential equation (\ref{presenceofcoll}), using the same initial conditions adopted in \cite{exp1}, that is, $f_g(z_{fin}, k)=1$, with the final redshift being equal to $z_{fin}=10$. Before presenting the results, we have to determine the allowed values of the wavenumber $k$ in order we remain within the sub-horizon approximation. From relation (\ref{subhorapprx}), using the present epoch's values for the scale factor and the Hubble parameter, we get that the wavenumber has to be $k>0.000156$, while the corresponding condition for non-collisional matter is $k>0.0001174$. Notice that we used the expression for the Hubble parameter given in relation (\ref{hubblepar}), with $g(a)$ being equal to,
\begin{equation}\label{ga}
g(a)=a^{-3}\Big{[} 1+\Pi_0+3w \ln (a) \Big{ ]}
\end{equation}
corresponding to collisional matter and with,
\begin{equation}\label{ga1}
g(a)=a^{-3}
\end{equation}
corresponding to non-collisional matter. In addition, we assume that only matter is present, thus disregarding radiation, since matter perturbations correspond to the matter domination era. In Fig. \ref{fig:oikonomouplots1} we present the plots of the growth factor $f_g(z)$ as a function of $z$, for collisional (blue line) and for non-collisional matter (red line) for $k=0.1$Mpc$^{-1}$ and $w=0.7$. 
As it can be seen, the collisional matter generates matter perturbations (blue line) that differ significantly, in reference to non-collisional matter perturbations (red line). This behavior is to be contrasted to the behavior of the parameters that describe the cosmological evolution of the two different matter species, like for example the Hubble parameter, which case the evolution of collisional and non-collisional matter is almost identical (see left plot of Fig. \ref{hubcurv}). Before closing, we have to mention that there is another quantity that characterizes the growth of matter perturbations, related to the growth factor of matter perturbations $f_g(z)$ and also to the matter density parameter $\Omega_m(z)$, known as growth index $\gamma (z)$. The growth index is related to the aforementioned quantities as follows \cite{exp1},
\begin{equation}\label{growthindex}
f_g(z)=\Omega_m(z)^{\gamma (z)}
\end{equation}
The growth index cannot be observed directly but can be determined from the observational data of both $\Omega_m(z)$ and $f_g(z)$. For a detailed analysis of the growth index for the exponential model (\ref{expmodnocurvcorr}), see reference \cite{exp1}.

\section{Analysis of the Results}

In order to have a global picture of the effects of collisional matter to the cosmological evolution of the universe at all stages, in this section we thoroughly discuss the results of the analysis we performed in this paper and combine these to the results obtained in \cite{oikonomoukaragiannakis}. In that paper we performed an analysis on the late time cosmological evolution of the $F(R)$ universe that contains collisional matter and compared the results to the $\Lambda$CDM model and also to $F(R)$ models that contain ordinary matter. As we explicitly demonstrated, the results were strongly model dependent and for some models, the evolution of the collisional matter $F(R)$ theory provided better fitting to the $\Lambda$CDM model, in comparison to the non-collisional $F(R)$ models. Therefore, the need for a deeper analysis of various aspects of cosmological evolution was necessary. In this paper we studied different aspects of cosmological evolution, and specifically we studied the dark energy oscillations and also the matter perturbations in a matter dominated universe. We used a popular exponential $F(R)$ model, with curvature corrections, which is given in relation (\ref{expmodnocurvcorr}). The resulting picture is that, in general, collisional matter generates a pronounced oscillatory behavior for dark energy, in reference to non-collisional matter (see Figures \ref{ycomp} and \ref{omegadecomp}). However, the cosmological evolution of the universe quantified in terms of the Hubble parameter, the effective equation of state parameter and the scalar curvature, looks quite similar in both collisional and non-collisional matter filled universes (see Figures \ref{hubcurv} and \ref{weffcomp}). In addition to this, by looking Table 1, we can see that some observational data coming from collisional matter filled universe, always for the $F(R)$ model under study, are quite closer to the observational data, in comparison to the non-collisional matter case. For example the value of the dark energy equation of state for $w=0.7$ in the collisional matter case, is much closer to the observed value. However, in the same case, the dark energy density for the non-collisional matter is closer to the observed value, than the corresponding value of the collisional matter case. In addition, collisional matter generates different matter perturbations, in comparison to the ones generated from non-collisional matter. We have to note that the same conclusions hold true for other $F(R)$ models we tried, for example the curvature corrected $R^2$-model, but we omit the details for brevity. Notice however that the present exponential model combines many elegant characteristics which render it as an ideal viable reference model for these studies.  

In conclusion, it is rather difficult to decide whether the effect of collisional matter makes the resulting picture better or worse. It is necessary to further scrutinize the effects of collisional matter so we can have more arguments in favor or against collisional matter. Some aspects of the cosmological evolution we did not address in this article, but should be thoroughly studied are the following: Firstly, one should perform a detailed analysis of the density perturbations in the framework of $F(R)$ theories. Note that density perturbations may be extracted from any first order tensor orthogonal to the four velocity, and describes the scalar part of any perturbation variables. There exist very stringent constraints coming from baryon acoustic oscillations and these could shed some light on whether we should consider the collisional matter a necessary ingredient of the matter content of our universe. Secondly, since we only restricted our study to $F(R)$ theories in the Jordan frame, one should try to study the evolution of a universe filled with collisional matter in the Einstein frame, by making a conformal transformation of the original $F(R)$ theory. Moreover, in order to be detached from the $F(R)$ framework, the same study we performed here should be done for scalar tensor theories, thus for gravitational theories that contain scalar fields. In the same study, it is possible to examine how the parametrization of the dark energy transition to the phantom epoch is affected by the presence of collisional matter. 
 
\begin{figure}[h]
\centering
\includegraphics[width=15pc]{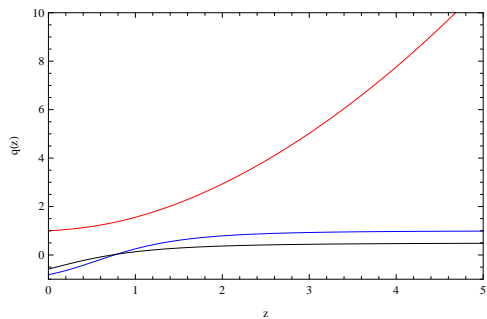}
\caption{Plots of $q(z)$ over z, for the model
$F(R)=R-2\Lambda \left ( 1-e^{\frac{R}{b\Lambda}} \right )-\tilde{\gamma}\Lambda \left (\frac{R}{3\tilde{m}^2} \right )^{1/3}$. The blue, red,
and black lines refer to collisional, non-collisional, and
$\Lambda\mathrm{CDM}$ models, respectively.}\label{myti}
\end{figure}

It worths to briefly present the late time cosmological evolution of the exponential model (\ref{expmodnocurv}) in terms of the deceleration parameter $q(z)$, as a function of the redshift $z$, a study which was absent in our previous work \cite{oikonomoukaragiannakis}. Following the approach we adopted in \cite{oikonomoukaragiannakis}, and omitting the details for brevity, in Fig. \ref{myti} we plot the deceleration parameter $q(z)$ for collisional matter (blue), non collisional matter (red) and for the $\Lambda$CDM model (black). As it is obvious, in this case too, the collisional matter filled universe gives better fit to the $\Lambda$CDM curve.

Before we close this section we comment on another rather interesting feature of the collisional matter filled universe. It is known that the cosmographic indication suggest that the cosmological standard model expressed in terms of the Hubble parameter, is extended in the following way \cite{caponew},
\begin{equation}\label{capowner}
H(z)\sim \sqrt{\Omega_m(1+z)^3+\ln (\alpha+\beta z)}
\end{equation}
Notice relation (\ref{fgd}), which when expressed in terms of the redshift $z$, takes the following form,
\begin{equation}\label{hubbleparcaponew}
H(z)=\sqrt{\tilde{m}^2y_H(z)+(1+z)^3\Big{[} 1+\Pi_0+3w \ln (1+z) \Big{ ]}+\chi (z+1)^{4}}
\end{equation} 
We have to notice the resemblance between the two Hubble parameters. In a future work, it worths trying to fit the collisional matter Hubble parameter (\ref{hubbleparcaponew}) with the cosmographically indicated one (\ref{capowner}).

\section*{Concluding Remarks}

We have studied the effect of collisional matter on the cosmological evolution of an viable curvature corrected exponential $F(R)$ model and compared the results to those coming from the same theoretical framework but with non-collisional matter. Particularly, we focused our study on the oscillatory behavior of dark energy during the matter domination era and at later epochs, an effect that is owing to the existence of diverging higher derivatives of the Hubble parameter. This phenomenon is present in the exponential $F(R)$ model, even in the presence of ordinary, that is, non-collisional matter, and as we explicitly demonstrated, this also occurs in the collisional matter case. Moreover, this oscillatory behavior is even more pronounced in the collisional matter case, and as the collisional matter equation of state parameter $w$ takes lower values, this phenomenon is not so intense. Specifically, this oscillatory behavior can be seen in the study of the dark energy equation of state parameter and for $z$ taking values in the matter domination era. However, one important result of our analysis is that the effective equation of state parameter oscillates but never crosses the phantom divide and in addition the oscillations are not so severe.

The cosmological evolution of collisional and non-collisional matter filled universes are almost indistinguishable, a fact that motivated us to study matter perturbations in order to reveal differences in the process of cosmic expansion. As we explicitly showed by studying the growth factor, the evolution of matter perturbations is different in the two forms of matter, so this can distinguish the cosmic evolution of collisional and non-collisional matter. The final picture is rather vague, since in some cases collisional matter produces a cosmology that fits better the observational data in comparison to non-collisional matter, always in the context of viable $F(R)$ theories. On the other hand, the oscillatory behavior of dark energy is pronounced in the collisional matter case, and as a result the final picture is not clear if the existence of some sort of interaction between ordinary non-relativistic matter can be considered as a true fact. Therefore, one should further scrutinize the effects of collisional matter, until a conclusion is reached. In addition, one should try to embed the collisional matter effects into other cosmological theoretical frameworks, such as scalar-tensor theories and also further study the evolution of matter perturbations more concretely. For an interesting work with respect to the latter see \cite{alvaro}. We hope to address these issues in a future work.


\begin{thebibliography}{}






\bibitem{riess} A.G. Riess et al. (High-z Supernova Search Team), Astronom. J. 116, 1009 (1998) [arXiv:astro-ph/9805201]

\bibitem{planck} P.A.R. Ade et al. [arXiv:1302.5082]

\bibitem{bicep} P. A. R. Ade et al. [arXiv:1403.3985], (2014)

\bibitem{mukhanov} V. Mukhanov, Physical foundations of cosmology, Cambridge, UK: Univ. Pr. (2005) 421 p; D. S. Gorbunov, V. A. Rubakov, Introduction to the theory of the early universe: Cosmological perturbations and inflationary theory, Hackensack, USA, World Scientific (2011) 489 p

\bibitem{kinezosvergados} Yeuk-Kwan E. Cheung, J.D. Vergados, arXiv:1410.5710

\bibitem{oikonomouvergados}V.K. Oikonomou, J.D. Vergados, Ch.C. Moustakidis, Nucl.Phys. B773 (2007) 19 [hep-ph/0612293]


\bibitem{reviews1} S. Nojiri, S. D. Odintsov, Int.J.Geom.Meth.Mod.Phys. 11 (2014) 1460006 [arXiv:1306.4426]; Int. J. Geom. Meth. Mod.Phys. 4 (2007) 115 [hep-th/0601213]
 
\bibitem{reviews2} S. Capozziello, V. Faraoni, Beyond Einstein Gravity, Springer, Berlin 2010


\bibitem{reviews3} F. S. N. Lobo, Dark Energy-Current Advances and Ideas, 173-204 (2009) [arXiv:0807.1640]


\bibitem{reviews4} S. Nojiri, S. D. Odintsov,  Phys.Rept. 505 (2011) 59 [arXiv:1011.0544]


\bibitem{reviews5} S. Capozziello, M. De Laurentis, Phys.Rept. 509 (2011) 167 [arXiv:1108.6266]


\bibitem{reviews8} K. Bamba, S. Nojiri, S. D. Odintsov, JCAP 0810 (2008) 045 [arXiv:0807.2575]

\bibitem{reviews9} S. Nojiri, S. D. Odintsov, Phys.Lett. B657 (2007) 238 [arXiv:0707.1941]

\bibitem{importantpapers1} S. Capozziello, S. Nojiri, S.D. Odintsov, A. Troisi, Phys.Lett. B639 (2006) 135 [astro-ph/0604431]; S. Nojiri, S. D. Odintsov, Phys.Rev. D77 (2008) 026007 [arXiv:0710.1738]

\bibitem{importantpapers2} S. Capozziello, V.F. Cardone, S. Carloni, A. Troisi, Int.J.Mod.Phys. D12 (2003) 1969 [astro-ph/0307018]

\bibitem{importantpapers3} S. Nojiri, S. D. Odintsov, Phys.Rev. D74 (2006) 086005 [hep-th/0608008]; A. de la Cruz-Dombriz, A. Dobado, Phys.Rev. D74 (2006) 087501 [gr-qc/0607118] 

\bibitem{importantpapers4} W. Hu, I. Sawicki, Phys.Rev.D76 (2007) 064004 [arXiv:0705.1158] 

\bibitem{importantpapers5} S. M. Carroll, V. Duvvuri, M. Trodden, M. S. Turner, Phys.Rev. D70 (2004) 043528 [astro-ph/0306438]; S. Capozziello, Int.J.Mod.Phys.D11, 483 (2002) [gr-qc/0201033]

\bibitem{importantpapers6} R. Myrzakulov, L. Sebastiani, S. Zerbini, Int.J.Mod.Phys. D22 (2013) 1330017 [arXiv:1302.4646]


\bibitem{importantpapers8} O. Bertolami, R. Rosenfeld, Int.J.Mod.Phys. A23 (2008) 4817 [arXiv:0708.1784]

\bibitem{importantpapers9} A. Capolupo, S. Capozziello, G. Vitiello, Int.J.Mod.Phys. A23 (2008) 4979 [arXiv:0705.0319]

\bibitem{importantpapers10}   P. K.S. Dunsby, E. Elizalde, R. Goswami, S. Odintsov, D. S. Gomez, Phys.Rev. D82 (2010) 023519 [arXiv:1005.2205]

\bibitem{importantpapers11} G. Cognola, E. Elizalde, S. Nojiri, S.D. Odintsov, L. Sebastiani, S. Zerbini, Phys.Rev. D77 (2008) 046009 [arXiv:0712.4017 ]


\bibitem{importantpapers11a} K. Bamba, Chao-Qiang Geng, Chung-Chi Lee, JCAP 1008 (2010) 021 [arXiv:1005.4574]


\bibitem{importantpapers12}   S. Nojiri, S. D. Odintsov, D. Saez-Gomez, Phys.Lett. B681 (2009) 74 [arXiv:0908.1269]  

\bibitem{importantpapers13} S. Capozziello, V. F. Cardone, A. Troisi, Phys.Rev. D71 (2005) 043503 [astro-ph/0501426]

\bibitem{importantpapers14} J. C.C. de Souza, Valerio Faraoni, Class.Quant.Grav. 24 (2007) 3637 [arXiv:0706.1223]; V. Faraoni, Phys.Rev. D74 (2006) 104017 [astro-ph/0610734];G. J. Olmo, Phys.Rev.Lett. 95 (2005) 261102 [gr-qc/0505101]; G. J. Olmo, Phys.Rev. D75 (2007) 023511 [gr-qc/0612047] 

\bibitem{importantpapers15} S. A. Appleby, R. A. Battye, A. A. Starobinsky, JCAP 1006 (2010) 005 [arXiv:0909.1737]
 

\bibitem{importantpapers17} S. A. Appleby, R. A. Battye, Phys.Lett.B654 (2007) 7 [arXiv:0705.3199]; S. A. Appleby, R. A. Battye, JCAP 0805 (2008) 019 [arXiv:0803.1081]

\bibitem{importantpapers18} A. Silvestri, M. Trodden, Rept. Prog. Phys. 72 (2009) 096901 [arXiv:0904.0024]

\bibitem{importantpapers19} E. Elizalde, E.O. Pozdeeva, S.Yu. Vernov, Phys.Rev. D85 (2012) 044002 [arXiv:1110.5806]

\bibitem{importantpapers20} V. Faraoni,  Phys.Rev. D75 (2007) 067302 [gr-qc/0703044]


\bibitem{sergeinojirimodel} S. Nojiri, S. D. Odintsov, Phys.Rev. D68 (2003) 123512 [hep-th/0307288]


\bibitem{capo} M. Sami, Curr. Sci. 97,887(2009) [arXiv:0904.3445]; Yi-Fu Cai, E. N. Saridakis, M. R. Setare, Jun-Qing Xia, Phys.Rept. 493 (2010) 1 [ arXiv:0909.2776]

 \bibitem{capo1} T. Padmanabhan, Phys.Rept. 380 (2003) 235 [hep-th/0212290]; K. Bamba, S. Capozziello, S. Nojiri, S. D. Odintsov, Astrophys. Space Sci. 342, 155 (2012) [arXiv:1205.3421]

\bibitem{peebles} P.J.E. Peebles, Bharat Ratra, Rev.Mod.Phys. 75 (2003) 559 [astro-ph/0207347]; V. Sahni, AIP Conf.Proc. 782 (2005) 166, J.Phys.Conf.Ser. 31 (2006) 115; M. Li, Xiao-Dong Li, S. Wang, Yi Wang, Commun.Theor.Phys. 56 (2011) 525 [arXiv:1103.5870]; A. Joyce, B. Jain, J. Khoury, M. Trodden [arXiv:1407.0059]

\bibitem{faraonquin} V. Faraoni, Int.J.Mod.Phys. D11 (2002) 471 [astro-ph/0110067]; V.K. Onemli, R.P. Woodard, Class.Quant.Grav. 19 (2002) 4607 [gr-qc/0204065] 

\bibitem{tsujiintjd} A. Gomez-Valent, J. Sola, S. Basilakos, arXiv:1409.7048; S. Basilakos, S. Nesseris, L. Perivolaropoulos, Phys.Rev. D87 (2013) 12, 123529; J. Khoury, A. Weltman, Phys. Rev. D69, 044026 (2004) [astro-ph/0309411 ]
 
\bibitem{quintense} Md. Wali Hossain, R. Myrzakulov, M. Sami, E. N. Saridakis, Phys.Rev. D89 (2014) 123513
 
\bibitem{saridakismyrzakulov} Md. Wali Hossain, R. Myrzakulov, M. Sami, Emmanuel N. Saridakis, arXiv:1410.6100 

\bibitem{oikonomoukaragiannakis} V.K. Oikonomou, N. Karagiannakis, arXiv:1408.5353

\bibitem{kleidis} K. Kleidis, N. K. Spyrou, Astron.Astrophys. 529 (2011) A26

\bibitem{fock} V. Fock, The theory of space, time and gravitation, Pergamon Press, 1959, London UK



\bibitem{exp0}  E. Elizalde, S.D. Odintsov, L. Sebastiani, S. Zerbini, Eur.Phys.J. C72 (2012) 1843 [arXiv:1108.6184]

\bibitem{exp1} K. Bamba, A. Lopez-Revelles, R. Myrzakulov, S.D. Odintsov, L. Sebastiani, Class.Quant.Grav. 30 (2013) 015008 [arXiv:1207.1009]

\bibitem{exp2} E. V. Linder, Phys.Rev. D80 (2009) 123528 [arXiv:0905.2962]

\bibitem{exp3} K. Bamba, A. Lopez-Revelles, R. Myrzakulov, S.D. Odintsov, L. Sebastiani, TSPU Bulletin 13 (2012) 128, 22-27 [arXiv:1301.3049]

\bibitem{exp4} V.K. Oikonomou, Gen.Rel.Grav. 45 (2013) 2467 [arXiv:1304.4089 ]

\bibitem{matsumoto} J. Matsumoto, Phys.Rev.D83 (2011) 124040 [arXiv:1105.1419]

\bibitem{caponew} S. Capozziello, M. De Laurentis, O. Luongo, arXiv:1411.2822

\bibitem{alvaro} A. de la Cruz-Dombriz, A. Dobado, A. L. Maroto, Phys.Rev.D77 (2008) 123515 [arXiv:0802.2999]





\end{thebibliography}
\end{document}